# Boron Fullerenes: A First-Principles Study

Nevill Gonzalez Szwacki*

*Physics Department, Texas Tech University, Lubbock TX 79409-1051, USA*
[*]e-mail: *nevill.gonzalez@ttu.edu*

A family of unusually stable boron cages was identified and examined using first-principles local density functional method. The structure of the fullerenes is similar to that of the $B_{12}$ icosahedron and consists of six crossing double-rings. The energetically most stable fullerene is made up of 180 boron atoms. A connection between the fullerene family and its precursors, boron sheets, is made. We show that the most stable boron sheets are not necessarily precursors of very stable boron cages. Our finding is a step forward in the understanding of the structure of the recently produced boron nanotubes.

**1. Introduction**

The chemistry of boron resembles that of carbon in its ability to *catenate* and form molecular networks. Unlike carbon, bulk boron cannot be found in nature and all known boron allotropes where obtained in the laboratory. All of them are based on different arrangements of $B_{12}$ icosahedrons. It is very natural to believe that not only carbon but also boron posses molecular allotropes (fullerenes and nanotubes). Experimental and theoretical research on the chemistry of boron nanomaterials is developing rapidly. The existence of quasi-planar[1] and tubular[2] boron clusters was predicted by theory and confirmed more recently by experiment.[3,4] Up to now, however, very little is known about the properties of these novel boron nanostructures.

In this work we will describe the properties of a family of boron nearly round cages which are built from crossing boron double-rings (DRs). The smallest members of the family, $B_{12}$ and $B_{80}$, where previously studied using first-principles methods.[5] Here we will show how to construct bigger cages with similar structural characteristics to those found in $B_{12}$ and $B_{80}$. We will also show the connection between boron cages and nanotubes as well as their precursors – boron sheets.

**2. Method**

The calculations were performed within the density functional theory framework, using ultrasoft Vanderbilt pseudopotentials[6] and local-density approximation for the Perdew-Burke-Ernzerhof exchange-correlation potential.[7] Computations were done using the plane-wave-based Quantum-ESPRESSO package.[8] The optimized geometries of the structures were found by allowing the full relaxation of the atoms in the cell until the atomic forces where less than $10^{-3}$ and $10^{-4}$ Ry/Bohr for the atomic cages and sheets, respectively. A proper *k*-point sampling for each system together with a 35 Ry cut-off for the plane-wave basis set have been used to ensure energy convergence to less than 1 meV/atom. To study properties of the fullerenes (nanotubes) the supercell geometry was taken to be a cubic (tetragonal) cell with sufficiently large lattice constant (constants) to avoid interactions between periodic replicas of the cluster. For infinitely long structures the supercell was optimized using variable cell optimization methods included in the program package.

**3. Boron fullerenes**

The unusual stability of the recently proposed fullerene of 80 boron atoms[5] and its structural similarities to the $B_{12}$ icosahedron motivates us to investigate larger cages with similar structural characteristics, that is, build from 6 crossing DRs but with larger diameters. It is known now that boron clusters with a number of atoms smaller than 20 are rather planar or quasi-planar and the $B_{12}$ icosahedron is energetically less favorable than the quasi-planar convex structure of $C_{3v}$ symmetry,[1,3,9] however, we would like to add formally the $B_{12}$ cage to the fullerene family as its smallest member. In Figure 1 we have shown what we call a next member of the fullerene family, which is made up of 180 atoms. Similarly to $B_{80}$, this cage posses $I_h$ symmetry but is more stable in energy than the $B_{80}$. As discussed previously[5] and also shown in Figure 1b we can clearly identify DRs as fragments of the $B_{180}$ round cage. For $B_{12}$, $B_{80}$, and $B_{180}$ the DRs have 10, 30, and 50 atoms, respectively. The only member of the family which is completely close is the $B_{12}$ icosahedron and the surface of the cage consists of 20 triangles. Since the fullerenes are built from a fixed number of DRs in larger cages the DRs cannot cover the whole sphere with a triangle network of atoms and the fullerenes will exhibit empty spaces – *holes*. All cages larger than $B_{12}$ will have 12 *holes*. The *holes* in $B_{80}$ are in the shape of pentagons, while the *holes* in $B_{180}$ are rather circular (more precisely they are closer to decagons; see e.g. Figure 1b). The size of the empty spaces increases with increasing cage diameter (see Figures 3a, 3b, and 3c).

The increase of the number of atoms by 20 between DRs belonging to two consecutive members of the fullerene family can be explained using the $B_{80}$ and $B_{180}$ cages as follows: each of the DRs in $B_{80}$ is adjacent to 10 pentagons (*holes*), so if, in order to obtain the $B_{180}$, we add one atom to each side of the pentagons the number of atoms of every DR increases by 10. In addition each DR requires still another 10 atoms in order to preserve its structure, so the total number of atoms for every DR increases by 20. We highlighted this in $B_{180}$ in Figure 1b where in one of the DRs the additional 20 atoms (respect to a DR in $B_{80}$) described above were colored in black and white.

The next in size fullerene after $B_{180}$ is made up of 300 atoms (it is built from DRs with 70 atoms). The optimized structure of this cage is shown in Figure 3c. The total number of atoms in $B_{12}$, $B_{80}$, and $B_{180}$ cannot be described by one general formula since the number of atoms shared by the DRs varies from one cage to another. Note that the black atoms belong only to one DR, in contrast, the white atoms are shared by two crossing DRs. However, for $B_{300}$ and all larger fullerenes the number of shared atoms is constant and the total number of atoms in the cage can be obtained using a simple formula $N(n) = N(B_{180}) + 120·(n - 3)$, where with $n$ ($n \geq 4$) we label the fullerenes starting from the smallest cage. The number of atoms in each of the $n$-fullerene DRs can be expressed by the formula $N_{DR}(n) = N_{DR}(B_{12}) + 20·(n - 1)$, where $n \geq 1$.

In Figure 2a we have plotted the cohesive energy ($E_{coh}$) of the four fullerenes discussed above versus the number of atoms in the cage. The less stable of them is the $B_{12}$ ($E_{coh}$ = 5.04 eV/atom) and the most stable is the $B_{180}$ ($E_{coh}$ = 5.77 eV/atom), which is 10 meV/atom more stable than the $B_{80}$. The $B_{300}$ cage has the same $E_{coh}$ as the $B_{80}$ fullerene to within 2 meV/atom. In the case of infinitely large cage $E_{coh}$ cannot be of course calculated exactly, however, it can be approximated by the $E_{coh}$ of an infinitely long stripe made up of boron atoms. Indeed, in a very large cage the 6 crossing DRs are almost isolated one from each other, and the atoms from the regions where the DRs cross give insignificant contribution to $E_{coh}$ in comparison to the rest of the atoms. The $E_{coh}$ = 5.69 eV/atom of the stripe will be then the lower limit for the energetic stability of large cages. In Figure 2a it is shown that the ($-E_{coh}$), after its minimum at $N$ = 180 atoms, increases for $B_{300}$ (by ~10 meV/atom), and that tendency should prevail also for larger cages until the lower limit for the $E_{coh}$ is reached. On the other hand it is interesting to note that the stability of DRs increase with increasing radius of the structures and the most stable is a DR with an infinite radius (stripe).[5] Therefore the $E_{coh}$ for the boron stripe is not only the lower limit for the energy of large cages but also the upper limit for the energetic stability of DRs.

The $B_{180}$ is not only the most stable (in energy) cage from those studied but also possesses almost perfect spherical shape. In Figure 2b we have plotted the radial distances, $r(\theta)$, of boron atoms belonging to $B_{80}$, $B_{180}$, and $B_{300}$ cages, from the center of each cage (center of mass), as a function of the spherical angle $\theta$. The average values for $r(\theta)$ are $R_1$ = 4.13, $R_2$ = 6.85, and $R_3$ = 9.39 Å for $B_{80}$, $B_{180}$, and $B_{300}$, respectively. The circles in Figure 2b represent the positions of boron atoms. In the case of $B_{80}$ there are 20 inner atoms and 60 outer atoms. The 60 outer atoms form a frame of the same structure as exhibited by the $C_{60}$ fullerene and lie on a sphere. The radius of this sphere is slightly larger than $R_1$. The 20 inner atoms are lying almost exactly at the centers of the hexagons of the $B_{60}$ frame at a radial distance of ~0.4 Å from the larger sphere. In the case of the $B_{180}$ fullerene the atoms are lying almost perfectly (to within 0.1 Å) on a sphere of radius $R_2$. More complex picture is present in the case of the $B_{300}$ fullerene. Here half of the atoms lie inside of a sphere of radius $R_3$ and half of them outside of this sphere. The more distant lie ~0.2 Å above or below the sphere surface. Should be pointed out that $B_{80}$ and $B_{300}$ exhibit braking symmetry distortions, which are however very small. A more detailed analysis has to be done to determine the nature of these structural distortions. In Table I we summarized our results for $E_{coh}$ and interatomic distances between neighboring boron atoms for boron fullerenes and also boron sheets (BSs) which are studied in the next section.

The structure of the cages influences also their electronic properties. For $B_{180}$ the highest occupied molecular orbital (HOMO) and lowest unoccupied molecular orbital (LUMO) are triply degenerate. (Similar result was reported previously for $B_{80}$.[5]) For $B_{300}$ the triple degeneracy both of HOMO and LUMO is slightly lifted (by less than 95 meV). This is most probably the results of the structural distortions mentioned above. The HOMO-LUMO energy gaps are 0.43 and 0.10 eV for $B_{180}$ and $B_{300}$, respectively. We have also calculated the electronic properties of the boron strip which was found to be metallic.

**4. Boron sheets**

To deepen the understanding of boron nanotubes and also boron cages it is important to know what is the structure of the BS. Several theoretical efforts,[10-15] using first-principles methods, have been done so far to understand the structure and properties of BSs. Most of these investigations have determined that the buckled triangle atomic lattice represents the most stable structure for the BS. However, it is intuitively understandable that the buckling of boron atoms is a response of the sheet to the internal stress imposed by the arrangement of the atoms in a triangle lattice. In principle it could not be discarded that there is an alternative arrangement in which the boron atoms may stay covalently bounded in a plane without buckling. Lau *et al.*[11] proposed that the BS can be formed by a network of triangle-square-triangle units of boron atoms. This structure, although planar, is less stable than the buckled triangle atomic lattice what was later on confirmed by the authors.[12] Very recently it was proposed a new BS which resembles the carbon honeycomb-like structure.[16] It is fully planar, possesses metallic properties, and is energetically more stable than the boron buckled triangle sheet. This structure is shown in Figure 3a (bottom) and will be labeled by us as $\alpha$-BS.

Our investigation of possible candidates for the BS we restricted to those which have structural similarities with the fullerenes studied above. In Figure 3 we have shown three cages, $B_{80}$, $B_{180}$, and $B_{300}$, and their corresponding sheets $\alpha$, $\beta$, and $\gamma$, respectively. In the top of Figures 3a, 3b, and 3c we have highlighted on each cage the characteristic atomic motif of the fullerene which also will appear on the sheet corresponding to it (see the sheets in the bottom of Figures 3a and 3b and the sheet in the left of Figure 3c). Let us forget for a moment about the *holes* in the cages, then the characteristic motif for the $B_{80}$ is a cluster of 7 atoms with one central atom lying almost in plain with a hexagonal chain of 6 atoms. This cluster has $C_{3v}$ symmetry. An isolated neutral cluster made up of 7 atoms has $C_{2v}$ symmetry.[9] The next two cages, $B_{180}$ and $B_{300}$, have motifs which are similar in shape and consist of quasi-planar structures of 12 and 18 atoms, respectively, and $C_{3v}$ symmetry. These clusters have 9 and 12 peripheral atoms and 3 and 6 central atoms in $B_{180}$ and $B_{300}$ cages, respectively. The interatomic distances between neighboring boron atoms in the fragment of $B_{180}$ are shown in Figure 1c. As it was mentioned in the previous section the isolated neutral cluster of 12 atoms is a quasi-planar convex structure of $C_{3v}$ symmetry. It was experimentally determined that this cluster has unusually large HOMO-LUMO gap of 2 eV.[3] It was also suggested that this cluster mast be extremely stable electrically and should be chemically inert.[9] Perhaps these unusual characteristics are responsible also for the outstanding stability of the $B_{180}$ fullerene.

All studied in this work BSs are fully planar and have metallic properties. We have found that the $\alpha$-BS which is a precursor of the $B_{80}$ cage is the most stable sheet over all studied (in agreement with recent findings[16]). The point symmetry of that sheet is $D_{6h}$. In Figure 3a (bottom) is shown the unit cell used for calculations of the $\alpha$-BS. The BSs $\beta$ and $\gamma$ corresponding to $B_{180}$ and $B_{300}$ fullerenes, respectively, have $D_{3h}$ symmetry. The shape of the *holes* in BSs will be determined by the type of atomic motif we are using to build the sheet. In the case of the $\alpha$-BS the *holes* are hexagons in the case of $\beta$ and $\gamma$ BSs the *holes* are distorted dodecagons and hexagons-like, respectively. It is important to observe that the BSs shown in Figure 3 can also be seen as built from staggered boron stripes. This observation may help to understand not only structural but also electronic properties of BSs.

The $E_{coh}$ for the $\alpha$-BS is 5.94 eV/atom and is bigger than the $E_{coh}$ for the $\beta$ and $\gamma$ BSs by 0.14 and 0.17 eV/atom, respectively. This is an interesting result since it means that the most stable structure for the BS does not necessarily have to be a precursor of very stable (in energy) boron cages.

Although there are some discrepancies between the $E_{coh}$ values obtained in this work (see Table I) and reported in the literature, the differences between $E_{coh}$ values corresponding to different sheets match very well. Indeed the $E_{coh}$ for the buckled triangle sheet is higher in energy than $E_{coh}$ for the flat triangle sheet by 0.24 eV/atom and this value is close to 0.21 and 0.22 eV/atom reported in Ref. 16 and Ref. 12, respectively. Similarly, $E_{coh}$ for the sheet $\alpha$ is higher in energy than $E_{coh}$ for the buckled sheet by 0.08 eV/atom, what represents slightly smaller value than 0.11 eV/atom obtained in previous calculations.[16]

## 5. Fullerene derived nanotubes

Carbon nanotubes and fullerenes are closely related structures. Capped nanotubes are elongated fullerene-like cylindrical tubes which are closed at the rounded ends.[2] To look for similar connections between boron nanostructures we have investigated boron nanotubes derived from the $B_{80}$ cages.

$B_{80}$-derived tubule can be obtained, in simplest way, by bisecting a $B_{80}$ molecule at the equator and joining the two resulting hemispheres with a cylindrical tube one monolayer thick and with the same diameter as the $B_{80}$. Almost all boron nanotubes studied theoretically so far are rolled up triangle sheets of boron atoms.[10, 13-15] It was natural then to take a cylindrical tube made of a triangle lattice as a candidate to join the two cups obtained from $B_{80}$. We have investigated only the simplest case when the fullerene is divided exactly at the middle of one of the DRs. The two hemispheres where then joined with a (15, 0) tube of "zigzag" geometry. However, upon relaxation this cupped nanotube significantly deformed. Since the $B_{80}$ cage can be seen as built from DRs we decided to design similarly the body of the nanotube as built of crossing stripes of double chains of boron atoms. Schematic of the resulting nanotube of 240 atoms is shown in Figure 4a. The two $B_{80}$ hemispheres are shown in red and the nanotube is shown in blue. After structural optimization (see Figure 4c) the nanostructure preserved its tubular form and the cups remained almost unchanged. It turns out that the body of this nanotube has a similar structure as the $\alpha$-BS.

The shortest nanotube is, of course, the $B_{80}$ cage and its $E_{coh}$ is 5.76 eV/atom.[5] We have also optimized the structures of two longer finite tubes – the $B_{160}$ (see Figure 4b) and the previously described $B_{240}$ (see Figure 4c). As expected the stability of the nanostructures increases with increasing lengths of the tube since $E_{coh}(B_{160}) = 5.81$ eV/atom and $E_{coh}(B_{240}) = 5.84$ eV/atom. The most stable is the infinite nanotube (see Figure 4d) with $E_{coh} = 5.87$ eV/atom. The HOMO-LUMO energy gaps for the $B_{160}$ and $B_{240}$ clusters are 0.33 and 0.02 eV, respectively, and the infinite tube was found to be metallic.

## 6. Summary

We are predicting the existence of a family of very stable boron fullerenes. The cages have similar structure consisting of 6 staggered boron DRs. The most stable fullerene is made up of 180 atoms and has almost perfect spherical shape. A recently proposed very stable BS of triangular and hexagonal motifs is a precursor of the $B_{80}$ cage. However, it was shown that the most stable sheets are not necessarily the precursors of very stable boron cages. Finally we have shown that the proposed fullerenes and novel boron nanotubes are closely related structures.

## Acknowledgements

The Interdisciplinary Centre for Mathematical and Computational Modelling of Warsaw University is thanked for a generous amount of CPU time.

## References

[1] I. Boustani, Phys. Rev. B 55, 16426 (1997).
[2] A. Gindulyte, W. N. Lipscomb, and L. Massa, Inorg. Chem. 37, 6544 (1998).
[3] H.-J. Zhai, B. Kiran, J. Li, and L.-S. Wang, Nat. Mater. 2, 827 (2003).
[4] D. Ciuparu, R. F. Klie, Y. Zhu, and L. Pfefferle, J. Phys. Chem. B 108, 3967 (2004).
[5] N. Gonzalez Szwacki, A. Sadrzadeh, and B. I. Yakobson, Phys. Rev. Lett. 98, 166804 (2007).
[6] D. Vanderbilt, Phys. Rev. B 41, 7892 (1990).
[7] J. P. Perdew, K. Burke, and M. Ernzerhof, Phys. Rev. Lett. 77, 3865 (1996).
[8] S. Baroni, *et al*., http://www.pwscf.org/.
[9] A. N. Alexandrovaa, A. I. Boldyreva, H.-J. Zhaib, and L.-S. Wangb, Coord. Chem. Rev. 250, 2811 (2006).
[10] J. Kunstmann and A. Quandt, Phys. Rev. B 74, 035413 (2006).
[11] K. C. Lau, R. Pati, R. Pandey, and A. C. Pineda, Chem. Phys. Lett. 418, 549 (2006).
[12] K. C. Lau and R. Pandey, J. Phys. Chem. C 111, 2906 (2007).
[13] J. Kunstmann, A. Quandt, and I. Boustani, Nanotechnology 18, 155703 (2007).
[14] I. Cabria, M. J. López, and J. A. Alonso, Nanotechnology 17, 778 (2006).
[15] M. H. Evans, J. D. Joannopoulos, and S. T. Pantelides, Phys. Rev. B 72, 045434 (2005).
[16] H. Tang and S. Ismail-Beigi, Phys. Rev. Lett. 99, 115501 (2007).

**Figure Captions**

**TABLE I.** Point symmetries, cohesive energies, and interatomic distances, $d_{BB}$, between neighboring boron atoms for fullerenes and sheets. For fullerenes and two sheets it is given the range for $d_{BB}$. In $\alpha$-BS, $d_1$ and $d_2$ are the distances between boron atoms in the triangular motif (each triangle has two $d_2$ sides and one $d_1$ side; the $d_1$ side is adjacent to the hexagonal motif). In the buckled triangle-sheet (TS), $h$ is the buckling height and $d_1$, and $d_2$ are the bond lengths. The values for the boron strip are given for comparison ($d_1$ and $d_2$ are the bond lengths).

**Fig. 1.** Top (a) and side (b) views of the optimized $B_{180}$ fullerene. It can be observed in (a) the almost perfect spherical shape of the cage. In (b) it is outlined a DR of 50 atoms. The black and white boron atoms represent additional 20 atoms with respect to the DR in $B_{80}$. Note that the black atoms are not shared by the DRs but each white atom is shared by two of them. The interatomic distances between neighboring boron atoms are shown in (c) using a fragment of the cage.

**Fig. 2.** (a) Cohesive energy per atom as a function of the number of atoms $N$ in the $B_N$ cluster. The horizontal line corresponds to the $E_{coh}$ = 5.69 eV/atom of the strip of boron atoms. The lines are a guide to the eye. (b) Radial distance, $r(\theta)$, of boron atoms belonging to $B_{80}$ (red circles), $B_{180}$ (blue circles), and $B_{300}$ (black circles) cages, from the center of mass of each cage, as a function of the spherical angle $\theta$. The red, blue, and black lines correspond to the average values of $r(\theta)$ for $B_{80}$, $B_{180}$, and $B_{300}$ cages, respectively.

**Fig. 3.** Three members of the fullerene family are shown: (a) $B_{80}$, (b) $B_{180}$, and (c) $B_{300}$. Each fullerene is accompanied by its precursor sheet. In all cages and sheets are highlighted (in blue) the corresponding atomic motifs which are discussed in the text. The unit cells used for calculations of sheets are shown in red.

**Fig. 4.** (a) Schematic of the $B_{240}$ nanotube described in the text. Optimized structures of the capped $B_{160}$ (b) and $B_{240}$ (c) nanotubes and the infinite (15, 0) nanotube (d). The clusters in (b) and (d) are shown in side (left) and front (right) views. The caps are two hemispheres of the $B_{80}$ molecule and the tubes are wrapped $\alpha$-BSs. For the infinite nanotube a unit cell of 80 atoms was used for calculations.

**TABLE I**

|  | symmetry | $E_{coh}$ (eV/atom) | $d_{BB}$ (Å) |
|---|---|---|---|
| $B_{80}$ | $I_h$ | 5.76 | 1.67 – 1.73 |
| $B_{180}$ | $I_h$ | 5.77 | 1.62 – 1.97 |
| $B_{300}$ | $I_h$ | 5.76 | 1.57 – 1.91 |
| $\alpha$ | $D_{6h}$ | 5.94 | $d_1 = 1.68, d_2 = 1.69$ |
| $\beta$ | $D_{3h}$ | 5.80 | 1.62 – 2.02 |
| $\gamma$ | $D_{3h}$ | 5.77 | 1.62 – 1.94 |
| flat TS | $D_{6h}$ | 5.62 | 1.71 |
| buckled TS | $C_{2v}$ | 5.85 | $d_1 = 1.62, d_2 = 1.86$ $h = 0.87$ |
| strip | $D_{2h}$ | 5.69 | $d_1 = 1.61, d_2 = 1.68$ |

**Fig. 1**

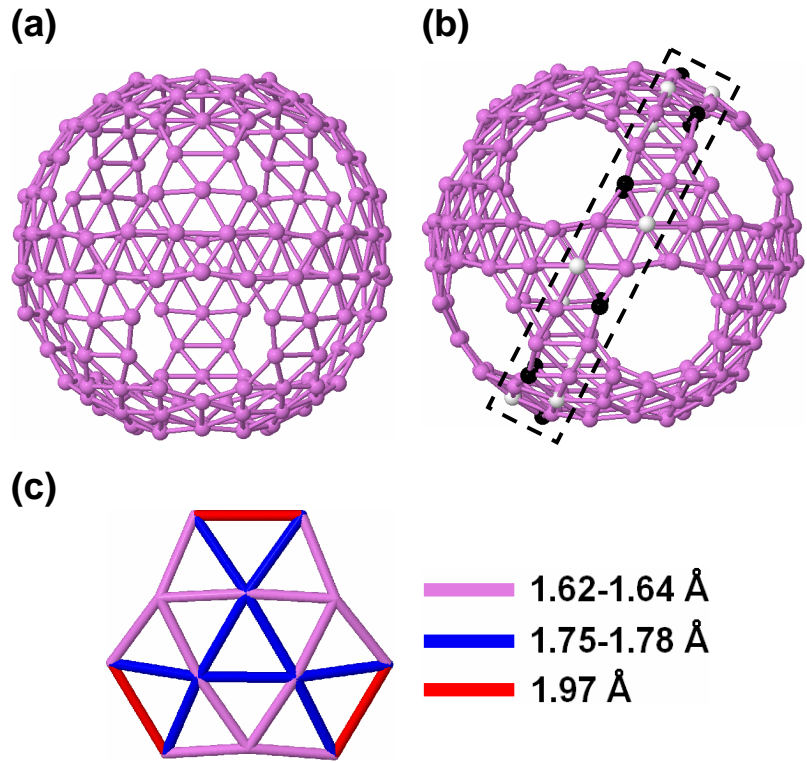

**Fig. 2**

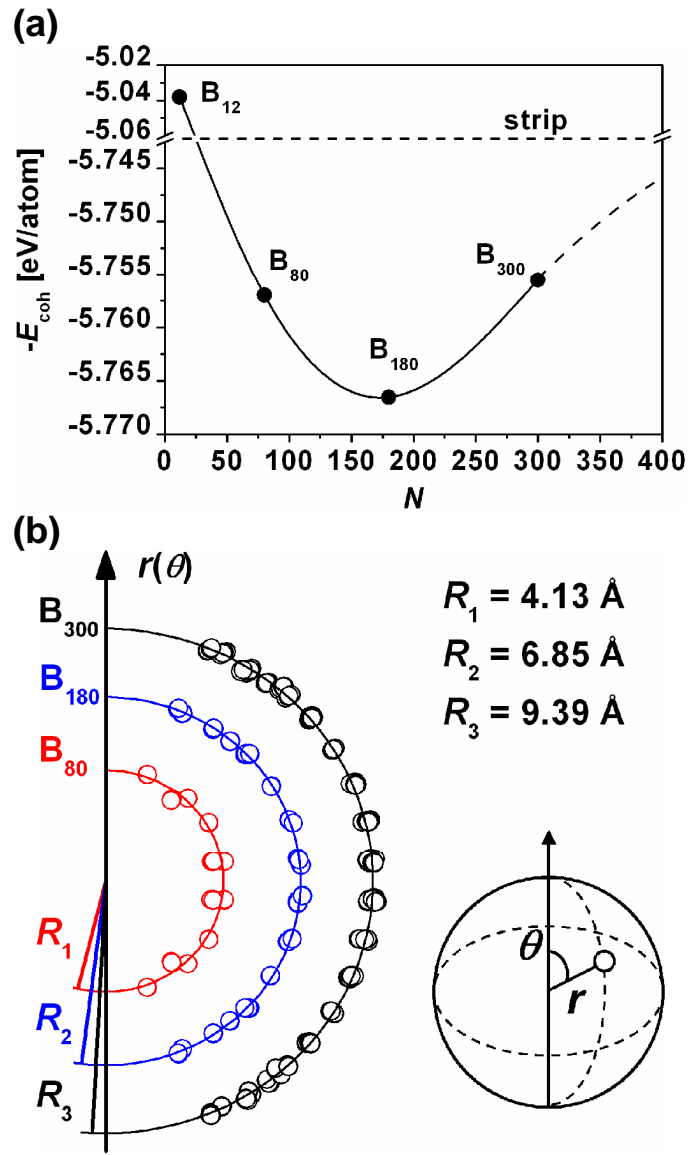

8 | Nevill Gonzalez Szwacki – *nevill.gonzalez@ttu.edu*

**Fig. 3**

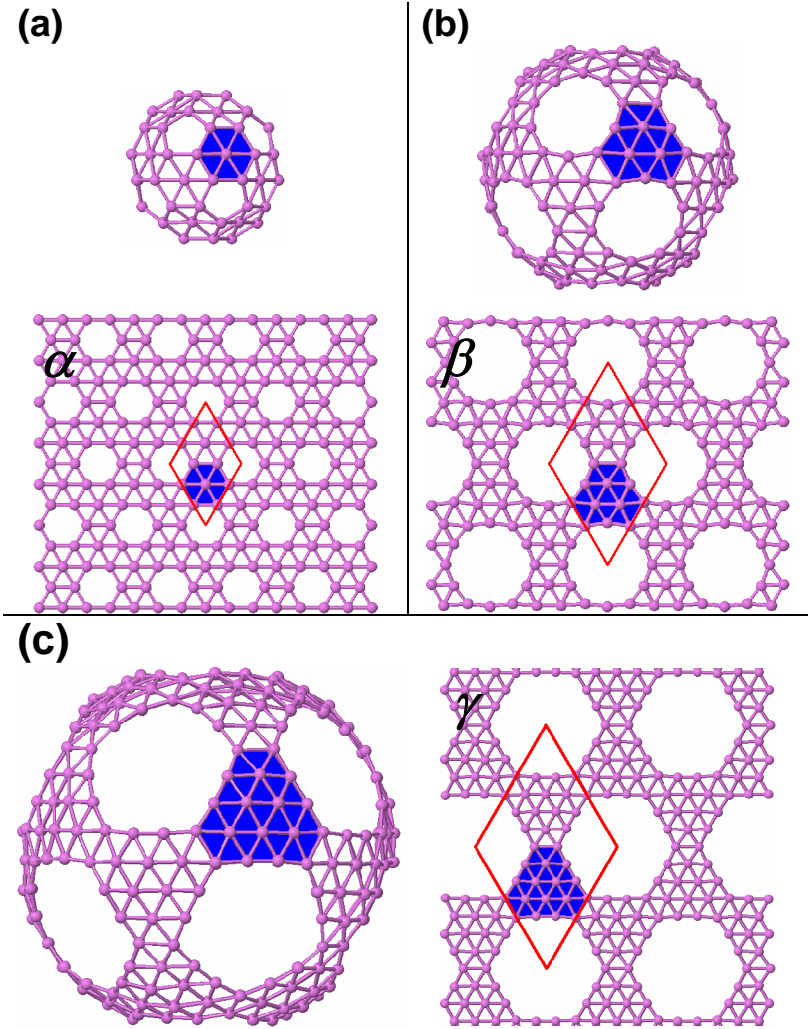

**Fig. 4**

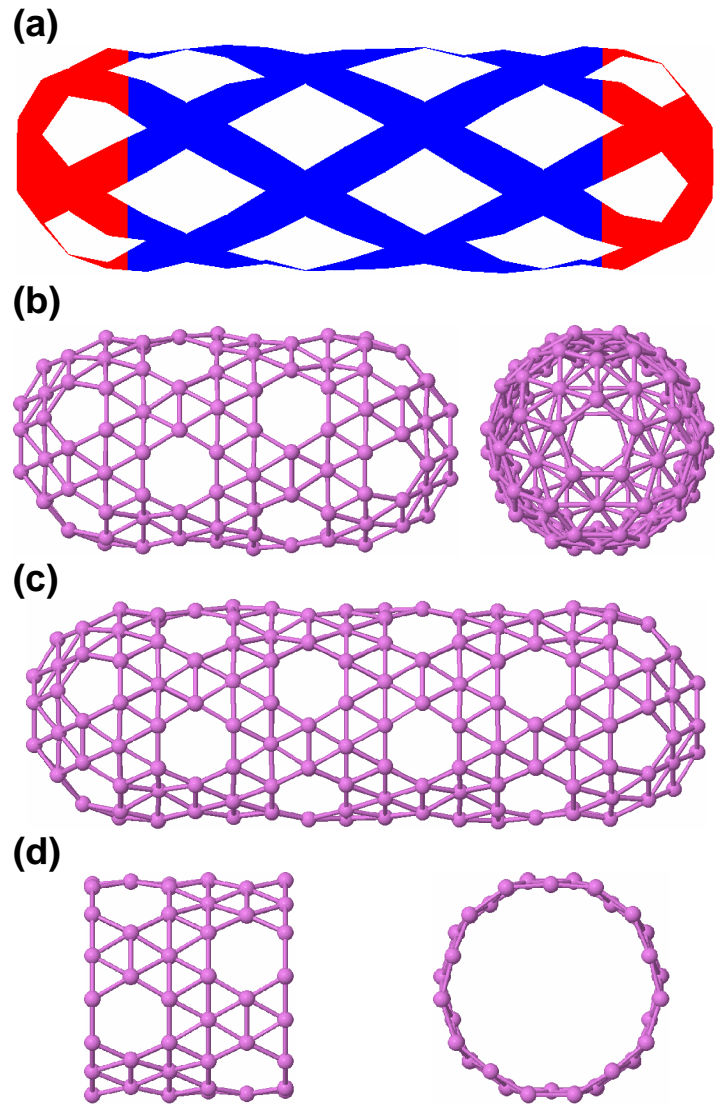